\pdfoutput=1

\documentclass[prl,twocolumn,showpacs,preprintnumbers,amsmath,amssymb,
superscriptaddress,nofootinbib]{revtex4-1}
\usepackage{epsfig}
\usepackage{graphicx}
\usepackage{color}

\newcommand{\la}{\langle}
\newcommand{\ra}{\rangle}
\newcommand{\be}{\begin{equation}}
\newcommand{\bea}{\begin{eqnarray}}
\newcommand{\ee}{\end{equation}}
\newcommand{\eea}{\end{eqnarray}}

\begin{document}

\preprint{CALT-TH-2017-074}

\title{Vector Effective Field Theories from Soft Limits}

\author{Clifford Cheung}
\affiliation{Walter Burke Institute for Theoretical Physics, California 
Institute of Technology, Pasadena, CA 91125}

\author{Karol Kampf}
\affiliation{Institute of Particle and Nuclear Physics, Charles
University, CZ 180 00 Prague}

\author{Jiri Novotny}
\affiliation{Institute of Particle and Nuclear Physics, Charles
University, CZ 180 00 Prague}

\author{Chia-Hsien Shen}
\affiliation{Bhaumik Institute for Theoretical Physics, University of California, Los Angeles, CA 90095}

\author{Jaroslav Trnka}
\affiliation{Center for Quantum Mathematics and Physics (QMAP), University of California, Davis, CA 95616}
\affiliation{Institute of Particle and Nuclear Physics, Charles
University, CZ 180 00 Prague}

\author{Congkao Wen}
\affiliation{Walter Burke Institute for Theoretical Physics, California 
Institute of Technology, Pasadena, CA 91125}
\affiliation{Bhaumik Institute for Theoretical Physics, University of California, Los Angeles, CA 90095}

%\date{\today}

\begin{abstract}
We present a bottom-up construction of vector effective field theories using the infrared structure of scattering amplitudes.   Our results employ two distinct probes of soft kinematics: multiple soft limits and single soft limits after dimensional reduction, applicable in four and general dimensions, respectively.  Both approaches uniquely specify the Born-Infeld (BI) model as the only theory of vectors completely fixed by certain infrared conditions which generalize the Adler zero for pions.  These soft properties imply new recursion relations for on-shell scattering amplitudes in BI theory and suggest the existence of a wider class of vector effective field theories.
\end{abstract}

\maketitle

\section{Introduction}

On-shell scattering amplitudes are fundamental physical observables in quantum field theory.  In recent years, these long-studied objects have spurred a multitude of exciting new developments: unexpected simplifications, hidden symmetries, and new mathematical structures completely invisible in the standard approach of Feynman diagrams. While most progress has centered on theories of maximal supersymmetry (SUSY) at high loop orders,  surprises have arisen even in the case of tree-level effective field theories (EFTs).

As is well-known, on-shell tree amplitudes in gauge theory and gravity are completely fixed by gauge invariance and proper factorization on poles,
\begin{equation}
\lim_{P^2 \rightarrow 0} A = \sum \frac{A_LA_R}{P^2},
\end{equation}
where the sum runs over all internal states. Alas, this approach does not uniquely specify EFTs, which exhibit higher-dimensional contact terms in the Lagrangian that are invisible on factorization kinematics.   This obstacle was overcome in \cite{Cheung:2015ota}, which showed how tree amplitudes in a broad class of scalar EFTs are completely fixed once factorization is supplemented by the additional physical criterion that the amplitude vanish as
\begin{equation}
\lim_{p \rightarrow 0} A = {\cal O}(p^\sigma),
\end{equation}
in the soft limit \cite{note:non-zero-soft}.
By constructing a general function that factorizes properly and {\it by assumption} conforms to certain values of $\sigma$, one discovers a remarkable class of {\it exceptional} theories: the non-linear sigma model (NLSM), Dirac-Born-Infeld (DBI) theory, and the special Galileon.  These theories exhibit soft behavior which is as strong as possible, exposing them as the scalar EFT analogs of gauge theory and gravity \cite{Cheung:2014dqa, Cheung:2016drk}.  

These scalar EFTs appear in a variety of disparate contexts, {\it e.g.}~in the Cachazo-He-Yuan scattering equations \cite{Cachazo:2013hca,Cachazo:2014xea,Cachazo:2016njl} as well as as certain worldsheet models \cite{Casali:2015vta}. Furthermore, they are mutually related by the Bern-Carrasco-Johansson double-copy construction \cite{Bern:2008qj,Cheung:2016prv} as well as the web of unifying relations for massless theories \cite{Cheung:2017ems,Cheung:2017yef}.

Notably, within this same orbit of topics appears ubiquitously a certain {\it vector} EFT: the Born-Infeld (BI) model.  This theory is a nonlinear extension of Maxwell theory which in $D$ dimensions has the Lagrangian
\begin{equation}
{\cal L}_{\rm BI} = 1-\sqrt{ (-1)^{D-1} {\rm det}(\eta_{\mu\nu}+ F_{\mu\nu})}  ,
\end{equation}
working in natural units with mostly minus metric convention. The purpose of this letter is to show that BI theory can also be uniquely specified by the infrared properties of its on-shell amplitudes.  Furthermore, the same methodology can be generalized to initiate an exploration of a larger class of vector EFTs.  Our results are built around two distinct soft probes which uniquely fix the BI action: a multiple chiral soft limit applicable to $D=4$ dimensions, and dimensional reduction to scalars applicable in any $D$.

To begin, consider a massless vector degree of freedom, which is described by a general Lagrangian expressed as a function of the gauge invariant Abelian field strength tensor, $F_{\mu\nu} = \partial_\mu A_\nu - \partial_\nu A_\mu$.  We employ a basis of scalar Lorentz invariants, $\la FF\dots F\ra$, where
\begin{align}
{\cal L} &= -\frac{1}{4}\la FF\ra + g_4^{(1)} \la FFFF\ra + g_4^{(2)} \la FF\ra^2  + g_6^{(1)} \la FF\ra^3\nonumber \\
&\hspace{0.5cm} +g_6^{(2)} \la FFFF\ra\la FF\ra + g_6^{(3)} \la FFFFFF\ra + \dots, \label{ansatz}
\end{align}
so $\la FF\ra = F_{\mu\nu}F^{\mu\nu}$, $\la FFFF\ra = F_{\mu\nu}F^{\rho\nu}F_{\rho\sigma}F^{\mu\sigma}$, etc., and all odd traces are identically zero. Note that imposing gauge invariance is not an additional assumption and simply encodes the existence of massless vector particles.  While this narrows the form of the Lagrangian ansatz we are still left with an infinite number of free coefficients, $g_n^{(m)}$.  From the above Lagrangian we then compute a tree-level $n$pt amplitude $A_n$ and fix the numerical coefficients $g_n^{(m)}$ by demanding certain soft properties of $A_n$. %In the case of scalar EFTs, soft behavior uniquely fixes the Lagrangians for NLSM, DBI and special Galileon.

\section{Uniqueness from Multi-Chiral Soft Limits}

First, let us focus on the case of $D=4$ where all possible interactions can be expressed in terms of two basic building blocks,
\begin{equation}
f = -\tfrac14 F_{\mu\nu} F^{\mu\nu} \qquad \textrm{and} \qquad  g= -\tfrac14 F_{\mu\nu} \widetilde F^{\mu\nu}  ,
\end{equation}
where $\widetilde F^{\mu\nu}  = \tfrac12 \varepsilon^{\mu\nu\rho\sigma} F_{\rho\sigma}$. Such an expansion is possible due to the Cayley-Hamilton relation for four-by-four matrices,
\begin{equation}
\langle F^n \rangle = -2 f \langle F^{n-2} \rangle + g^2 \langle F^{n-4}\rangle\,,
\end{equation}
using our earlier notation.   Assuming parity, we straightforwardly construct the most general effective Lagrangian for a massless vector particle,
\begin{equation}\label{eq:genL}
{\cal L} = f + a_1 f^2 + a_2 g^2 + b_1 f^3 + b_2 f g^2  +\ldots ,
\end{equation}
where $g$ enters only in even powers.
This Lagrangian covers an huge range of EFTs, including {\it e.g.}~the well-known Euler-Heisenberg theory describing quantum electrodynamics at low energy, as well as our target BI theory.

Next, let us consider the amplitudes corresponding to this general Lagrangian.  Starting at 4pt, there are three possible helicity configurations modulo helicity conjugation: $----$, $---+$, $--++$. In our conventions all particles are outgoing, so $+/-$ denote positive/negative-helicity particles, respectively. The 4pt amplitudes for the all-but-one same helicity configuration ($---+$) are zero.  However, we are still left with two independent on-shell amplitudes, which in spinor helicity variables are
\begin{align}
A_{----} &= \tfrac12 (a_1-a_2)(\la12\ra^2\la34\ra^2+{\rm perm}),\nonumber\\
A_{--++} &= \tfrac12 (a_1+a_2)\la12\ra^2[34]^2 .
\end{align}
For the moment, we make the assumption that the only non-vanishing  amplitudes are {\it helicity conserving}, {\it i.e.}, have equal numbers of positive- and negative-helicity particles, so $a_1=a_2$. This criterion alone does not fix the theory completely but it  will simplify our present analysis.  As we will see later on, helicity conservation can actually be dropped as an assumption in favor of a special infrared property of amplitudes.

With 4pt squared away we now compute the 6pt amplitude, $A_{---+++}$. Here we recycle the 4pt on-shell amplitude as a 4pt Feynman vertex, together contributions from a general 6pt contact term. However, as it turns out, the latter does not exist: due to considerations of little group weight and mass dimension, the only allowed contact term is $\langle 12\rangle\langle23\rangle\langle31\rangle[45][56][64]$, which vanishes identically after symmetrizing on (123) and (456).  Hence, the 6pt amplitude is given uniquely by factorization diagrams involving the 4pt vertex,

\begin{picture}(100,60)
\qquad\qquad\qquad\includegraphics[scale=0.6]{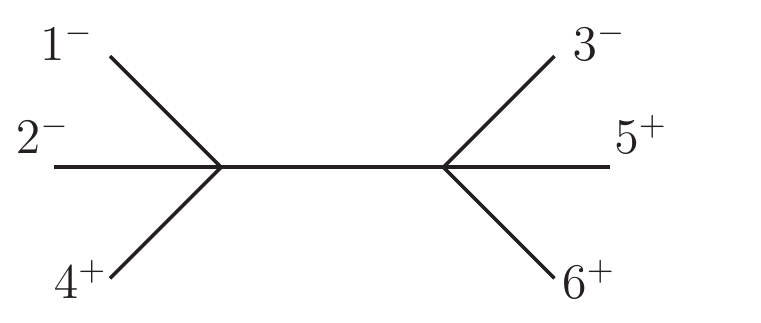}%\put(20,20){+ perm}
\end{picture}
\begin{align}
A_{---+++} &= \frac{\langle 12\rangle^2 [56]^2 \langle 3 | 1+2 | 4]^2}{s_{124}} + {\rm perm.,}
\end{align}
with permutations in the diagram tacitly assumed.  This amplitude scales as ${\cal O}(1)$ in the single soft limit, so it has not interesting in this respect.  However, we discover highly non-trivial infrared behavior if we take a {\it multi-chiral soft limit}, defined by sending  $\widetilde\lambda_1,\widetilde\lambda_2,\widetilde\lambda_3\rightarrow \epsilon$ {\it or\/} $\lambda_4,\lambda_5,\lambda_6\rightarrow \epsilon$ 
\bea
%\lim_{\overset{\!\scalebox{.7}{$\scriptscriptstyle(\sim)$}}{\lambda_{i}}\rightarrow \epsilon} \lim_{\overset{\!\scalebox{.7}{$\scriptscriptstyle(\sim)$}}{\lambda_{i}}\rightarrow \epsilon} A_{---+++} = {\cal O}(\epsilon) \,.
\lim_{\overset{\sim}{\lambda}_{-}\rightarrow \epsilon \;\, \text{or}\; \lambda_{+}\rightarrow \epsilon} \;  A_{---+++} = {\cal O}(\epsilon) \, ,
\eea 
where the $+/-$  subscripts on the spinors are shorthand for all legs of a given helicity.  Alternatively, we could instead send $\lambda_1,\lambda_2,\lambda_3\rightarrow \epsilon$ or $\widetilde\lambda_4,\widetilde\lambda_5,\widetilde\lambda_6\rightarrow \epsilon$ which gives analogous behavior ${\cal O}(\epsilon^7)$ with the extra $\epsilon^6$ suppression trivially entering through $\lambda$'s or $\tilde{\lambda}$'s in the polarization vectors. Interestingly, similar behavior can be achieved when only two of three spinors of given type are sent to zero. In this case, individual terms scale as ${\cal O}(1)$ so a cancelation must occur between diagrams. The crucial test of this approach is the 8pt amplitude given by the set of Feynman graphs:

\begin{picture}(100,140)
\includegraphics[scale=0.6]{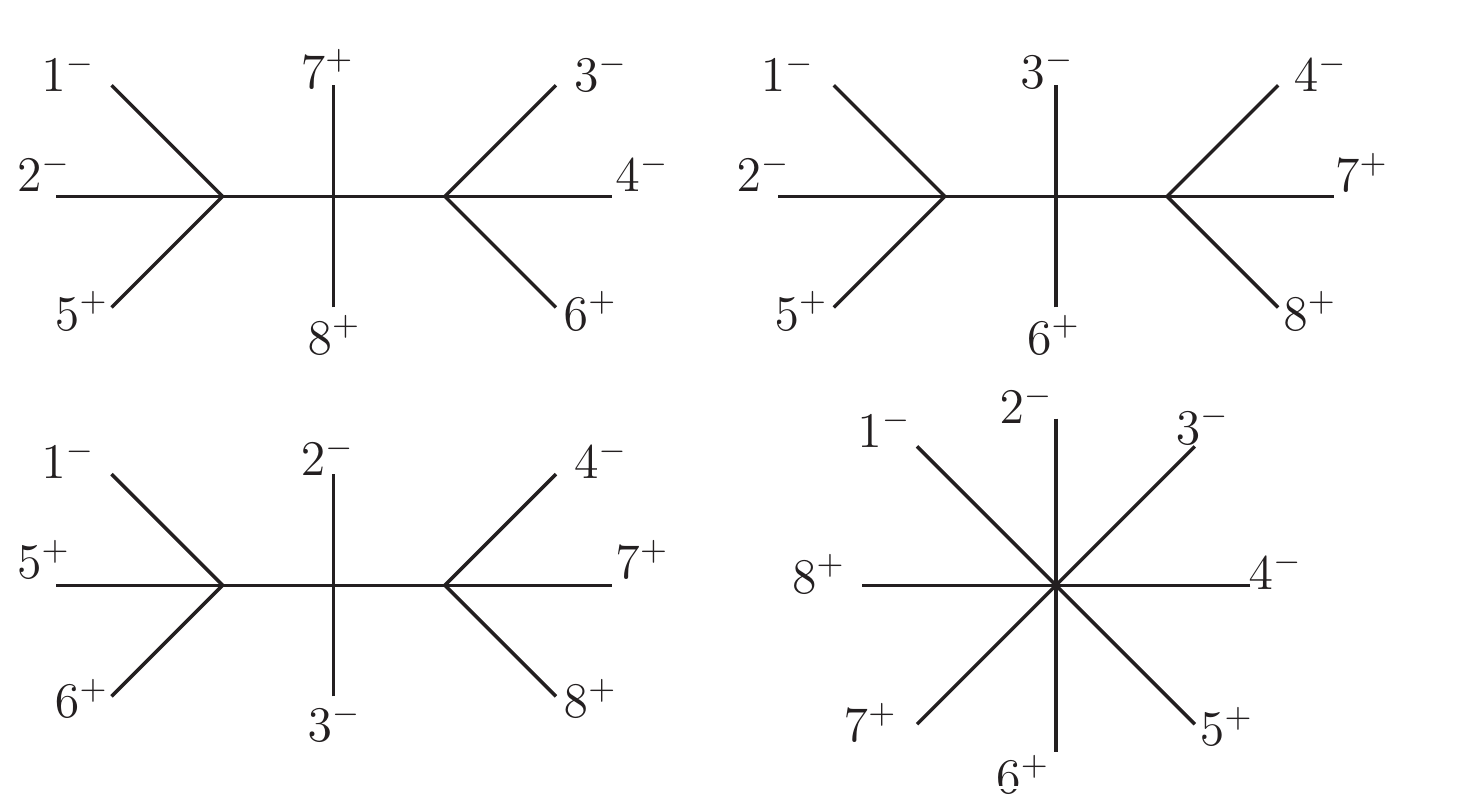}
\end{picture}

As discussed, the diagrams with 6pt vertices are absent but there is an 8pt contact term with an unfixed coefficient,
\begin{align}
A_{----++++} &= \frac{1}{2} \frac{ [ 5 | (1+2) (3+4) | 6]^2 \langle12\rangle^2\langle34\rangle^2 [78]^2}{s_{125}s_{346}} \notag\\
&+  \frac{ \langle 3 | 1+2| 5]^2 \langle4|7+8|6]^2 \langle12\rangle^2[78]^2 }{s_{125}s_{478}}\notag\\
&+ \frac{1}{2} \frac{ \langle 1 | (5+6) (7+8) | 4\rangle^2 
\langle23\rangle^2 [56]^2 [78]^2}{s_{156}s_{478}}\notag\\
&+ k\, \langle12\rangle^2
\langle34\rangle^2 [56]^2[78]^2 
+ \text{perm}\,.
\end{align}
As it turns this expression does not have any special behavior for single or double-chiral soft limit, but if we send $\widetilde\lambda_1,\widetilde\lambda_2,\widetilde\lambda_3,\widetilde\lambda_4\rightarrow\epsilon$ or $\lambda_5,\lambda_6,\lambda_7,\lambda_8\rightarrow\epsilon$, we again obtain vanishing behavior \cite{note:8ptsoft}  only if the coefficient of the contact term is set to $k=-1.$ Analogously, for 10pt amplitude there are no contact terms allowed so it is automatically ${\cal O}(\epsilon)$ behavior in the chiral soft limit when four or five appropriate spinors are set to zero. For 12pt amplitude there is a single contact term, $(\langle12\rangle^2\langle34\rangle^2\langle56\rangle^2[78]^2[9\,10]^2[11\,12]^2 +$ perm.), whose coefficient is fixed by demanding the ${\cal O}(\epsilon)$ behavior in the multi-chiral soft limit. This generalizes for any $n$: for each $n=4k$ there is a new contact term whose coefficient is uniquely fixed by appropriate multi-chiral soft limit. It is easy to see that translating back to Lagrangian, in $D=4$ this gives 
\begin{equation}
{\cal L}_{\rm BI} = 1 - \sqrt{1 - 2f - g^2} ,
\end{equation}
which is the action for BI theory. 

Interestingly, the initial assumption of helicity conservation can be dropped if we apply a generalization of the above multi-chiral soft behavior. For an amplitude with $n$ $-$ helicity and $m$ $+$ helicity with $n\leq m$, it is sufficient to require that
\begin{equation}
A(1^-2^-\ldots n^- (n+1)^+\ldots (n+m)^+) = {\cal O}(\epsilon)
\end{equation}
for the anti-holomorphic soft limit, $\widetilde{\lambda}_i \rightarrow \epsilon$ for $i=1\ldots n$. The case with $n=0$ must be trivially zero and the combinations of helicities with $n>m$ are obtained simply by the helicity conjugation.

Last but not least, it is possible to automate the physical criteria of factorization together with multi-chiral soft behavior by constructing an on-shell recursion relation where the spinors are shifted according to \cite{Cheung:2015cba,Cachazo:2016njl}
\begin{equation}\label{eq:shiftlambdas}
\widetilde{\lambda}_i \rightarrow \widetilde{\lambda}_i(1-z)\qquad \textrm{and} \qquad \lambda_k \rightarrow \lambda_k+z \eta_k ,
\end{equation}
for $i=1,\dots,\frac{n}{2}$ and $k=n-1,n$. The shift of $\widetilde{\lambda}_i$ probes the multi-chiral soft limit while shifting $\lambda_k$ to ensure momentum conservation provided
$$
\eta_{n-1} = -\frac{1}{[n-1\,n]}\sum_{i=1}^{n/2} [i\,n]\lambda_i,\,\, \eta_{n} = \frac{1}{[n-1\,n]}\sum_{i=1}^{n/2} [i\,n-1]\lambda_i .
$$
Because of the multi-chiral soft limit behavior, the amplitudes scale as $A_n = {\cal O}(1-z)$ for $z=1$ and as $A_n={\cal O}(1)$ for $z=\infty$, which can also be checked by the inspection of individual Feynman diagrams.  Applying the Cauchy formula to the shifted amplitude $A_n(z)$, we obtain
$$
\int \frac{dz\, A(z)}{z(1-z)} = 0,
$$
where the pole for $z=1$ in the denominator is canceled by the vanishing of $A(z)$. Summing over all other poles of $A(z)$ -- factorization channels -- gives us the recursion formula for $A_n$,
\begin{equation}
A_n = \sum_I \frac{A_L(z_{I_-})A_R(z_{I_-})}{P_I^2(1-z_{I_-}/z_{I_+})(1-z_{I_-})} + (z_{I_-}\leftrightarrow z_{I_+}),
\end{equation}
where the sum is over factorization channels $I$ and $z_{I_\pm}$ are roots of equation $\widehat{P}_I^2(z)=0$.

\section{Uniqueness from Supersymmetry}

The soft structue of the BI action can be derived using SUSY. As is well known, BI theory corresponds to the pure bosonic sector of the EFT describing spontaneous symmetry breaking of ${\cal N}=2$ to ${\cal N}=1$ SUSY \cite{Bagger:1996wp}. The full set of physical degrees of freedom are the BI photon $A_\mu$ and goldstino $\psi$.  The broken SUSY generators are realized through non-linear transformations. The goldstino transforms under non-linear SUSY according to generalized shift,
\begin{equation}
\psi_A \rightarrow \psi_A + \eta_A + \dots ,
\end{equation}
so amplitudes exhibit a vanishing ${\cal O}(p)$ soft limit for the goldstino when $p\rightarrow 0$. 

The unbroken ${\cal N}=1$ SUSY implies a Ward identity relating the pure photon amplitudes to the ones with two goldstinos,
\begin{align} \label{eq:SUSYWI}
&\tilde{\lambda}^{\dot{\alpha}}_1 A(1^-2^-\dots n/2^-(n/2+1)^+\dots n^+)\\
&=-\!\! \sum_{i=n/2+1}^n \tilde{\lambda}^{\dot{\alpha}}_i A(\psi_1^-2^-\dots n/2^-(n/2{+}1)^+\dots \psi_i^+ \dots n^+)\nonumber .
\end{align}
In the soft limit, defined by $\lambda_i\rightarrow 0$ for all $i > {n\over 2}$, the right-hand side is zero because the amplitude exhibits the goldstino soft zero.  Naively, there is the subtlety that the multi-chiral soft limit could induce a soft pole to cancel this Adler zero. However, such a pole does not appear because the factorization channel either vanishes by helicity conservation or is non-singular due to the specific form of the 4pt vertices. Thus, we conclude that the left-hand side of Eq.~(\ref{eq:SUSYWI}) vanishes, which is our conjectured soft theorem. Contracting both sides of Eq.~(\ref{eq:SUSYWI}) with $\tilde{\lambda}^{\dot{\alpha}}_j$ for any $j$ of a positive-helicity photon, we find that the BI amplitude also vanishes in the multi-chiral soft limit  $\lambda_i\rightarrow 0$ for $({n \over 2}-1)$ the positive-helicity photons, as we have also discussed in previous sections. It is the amplitude method which further tells that this property fixes the theory uniquely. 

\section{Uniqueness from Dimensional Reduction}

BI theory can also be fixed uniquely by a combination of soft limits and dimensional reduction.  In particular, we constraint a general amplitude for a massless vector  demanding that its dimensionally reduced amplitudes describe DBI scalars, whose dynamics are in turn completely specified by enhanced soft behavior.   Conveniently, dimensional reduction can be applied directly at the level of amplitudes.  To begin, consider an $n$pt amplitude in general $D$ dimensions, partitioning all  $n$ legs into $p$ sets, $\{ I_1|\dots| I_p\}$.  Here each set is interpreted as an extra dimension in which a subset of vectors are polarized, thus becoming scalars under dimensional reduction.  Since these extra-dimensional polarizations are orthogonal to the physical momenta, we set $(e_i\cdot p_j)=0$ and 
\begin{equation}
(e_i\cdot e_j)=\left\{ \begin{array}{cc}
1 & \!\!\!\! ,\quad i,j\in I_a \\
0 &,\quad \textrm{otherwise}
\end{array}
\right\} .
%1\,\,\mbox{for $i,j\in I_a$}, \quad (e_i\cdot e_j)=0\,\,\mbox{otherwise}
\end{equation}
The resulting dimensionally reduced amplitude describes $p$ flavors of scalar particles whose momenta are restricted to $(D-p)$ dimensions. Since the starting Lagrangian can be expressed in terms of field strengths, we know that the resulting scalars are derivatively coupled and trivially exhibit ${\cal O}(p)$ soft behavior. However, by demanding an enhanced ${\cal O}(p^2)$ soft limit on these scalars, these amplitudes are constrained to be scalar DBI amplitudes and we find that the original theory is uniquely BI theory.  

To describe a general dimensional reduction we introduce a notation $\lbrace a_1|a_2| \dots \rbrace$ where $a_i$ denotes number of photons reduced in a given extra dimension, {\it i.e.}, corresponding to the same set $I_{a_i}$. For example, starting from an $n$pt photonic amplitude in $D$ dimensions, we can reduce all photons to $\lbrace n\rbrace $ yielding a single scalar theory in $D-1$ dimensions, or reduce a subset of photons to $\lbrace a|b \rbrace $, yielding $a$ scalars of one flavor and $b$ scalars of another flavor propagating in $D-2$ dimensions, etc.

For concreteness, let us consider a 4pt example where the Lagrangian is a linear combination of two terms,
\begin{equation}
{\cal L}_{4}= c_1 \langle FFFF \rangle + c_2 \langle FF\ra^2 , 
\end{equation}
stipulating that dimension-reduced amplitude $\lbrace 4\rbrace $ has enhanced soft behavior, $A_4={\cal O}(p^2)$. This fixes the relative coefficients, $c_2=c_1/4$. 

Moving on to 6pt case, we take the general Lagrangian,
\begin{equation}
{\cal L}_{6}= d_1 \langle FFFFFF \rangle + d_2 \langle FFFF\ra\la FF\ra + d_3\la FF\ra^3.
\end{equation}
In this case, dimensional reduction to a single extra dimension $\lbrace 6\rbrace $ is insufficient to fix the Lagrangian.  The reason for this is that a certain linear combination of interaction terms actually vanishes when sending $(e_i\cdot p_j)=0$ and $(e_i\cdot e_j)=1$, and thus does not enter the soft constraint.  If, however, we also demand that the amplitude $A_6={\cal O}(p^2)$ for all possible dimensional-reductions into one $\lbrace 6\rbrace $, two $\lbrace 4|2\rbrace $, or three extra dimensions $\lbrace 2|2|2\rbrace $, we do obtain a unique solution. %We have also verified that the same property holds at 8pt.

That BI theory is uniquely fixed from its dimensionally reduced DBI amplitudes is actually obvious in hindsight.  In particular, any tree amplitude of vectors can be interpreted as polynomial in $(e_i \cdot e_j)$, with coefficients that depend on $(e_i \cdot p_j)$ and $(p_i \cdot p_j)$.  The dimensionally reduced amplitudes can be obtained from the original expression simply by applying derivatives with respect to $\partial / \partial (e_i \cdot e_j)$ \cite{Cheung:2017ems}.  Since the dimensionally reduced amplitudes are simply derivatives of the original amplitude, they uniquely fix the original expression up to a ``constant'' which depends only on $(e_i \cdot p_j)$ and $(p_i \cdot p_j)$.  However, any such term cannot itself be gauge invariant so it must be related to terms involving $(e_i \cdot e_j)$ which have already been fixed. 

An even more restricted operation also uniquely fixes the BI Lagrangian in Eq.~(\ref{ansatz}): reduce only a single pair of photons $e_i$, $e_j$ to scalars rather than all photons. In this case we set $(e_{i,j}\cdot p)=(e_{i,j}\cdot e_k)=0$ and $(e_i\cdot e_j)=1$, where $k$ denotes all other labels, yielding an amplitude of two scalars and $n-2$ photons. This is the limit $\lbrace 2\rbrace $. Demanding the soft limit behavior $\mathcal{O}(p^2)$ for either of the scalars also fixes the BI action which we have checked explicitly up to eight points.   This directly implies that the original vector amplitude can be expressed purely in terms of amplitudes involving two scalars, so
\bea \label{eq:reductionID}
A_n = \sum_{i<j} (e_i \cdot e_j) A(i,j)\big{|}_{(e\cdot e)^m \rightarrow {(e\cdot e)^m \over m}} \, ,
\eea
where each term of the form $(e\cdot e)^m$ is rescaled by a symmetry factor $1/m$ to eliminate overcounting and $A(i,j)$ is the amplitude with photons $i$ and $j$ dimensionally reduced to scalars~\cite{note:reductionID}. Importantly, since $A(i,j)$ has two DBI scalars it is uniquely fixed by its enhanced soft behavior, so it can be constructed using soft recursion \cite{Cheung:2015ota}.  So Eq.~(\ref{eq:reductionID}) is in turn a working definition of all tree amplitudes in BI theory.

\section{Vector galileon-like theories}

It is straightforward to generalize the construction of previous sections to a vector theory with even more derivatives. While BI theory has one derivative per field, the next interesting case corresponds to a Lagrangian of the schematic form,
\begin{equation}
{\cal L} = F^2 + \partial^2F^4 + \partial^4F^6 + \partial^6F^8 + \dots, \label{ansatz2}
\end{equation}
corresponding to the power counting of the scalar Galileon.  In detail, there are 3 terms of the form $\partial^2F^4$ and 64 terms of the form $\partial^4F^6$ in general $D$, modulo terms that can be eliminated by equations of motion.   The obvious extension of our previous results is to constrain Eq.~(\ref{ansatz2}) with a stronger ${\cal O}(\epsilon^3)$, in analogy with the soft behavior of the special Galileon. 

Notably, there is a no-go theorem forbidding vector particles with a Galileon symmetry \cite{Deffayet:2013tca} (see also an interesting recent discussion on SUSY Galileons \cite{Elvang:2017mdq}). However, this obstruction is evaded \cite{Deffayet:2010zh} if one considers multiple flavors of scalar Galileon or $p$-form Galileons for even $p$.  More importantly, in our case here we do not seek a theory with a bona fide Galileon symmetry but rather a theory of ``Galileon''-like interacting vectors with the same power counting as the scalar Galileon and similarly exceptional infrared properties. Here we offer partial evidence of the existence of a Galileon-like vector theory.

In $D=4$ we can construct an analog of Eq.~(\ref{eq:genL}) and impose more severe vanishing under chiral multi-soft limits. We again demand that only helicity conserving amplitudes are non-zero and for 4pt we get a single term
\begin{equation}
A_{--++} = \la12\ra^2[34]^2 s_{12} \label{higher4pt},
\end{equation}
while for 6pt we obtain five independent contact terms, in contrast to zero for BI power counting. Constructing the 6pt amplitude from factorization terms and contact term we find five free coefficients. By imposing the chiral multi-soft limit, $\lambda_4,\lambda_5,\lambda_6\rightarrow0$ we find ${\cal O}(\epsilon^4)$ for factorization term and one of the contact terms, with all other contact terms behaving worse. However, there is no choice of contact term coefficients that can accommodate an even stronger multi-chiral soft limit, so we cannot uniquely fix the amplitude from this procedure. This becomes even more dramatic at 8pt when one encounters contact terms with even stronger ${\cal O}(\epsilon^6)$ which are completely unfixed. While unsatisfactory in general we can still ask a less ambitious question: it is possible to start with Eq.~(\ref{higher4pt}) and consistently add contact terms such that we obtain an improved soft-limit behavior as a result of cancelation between individual Feynman diagrams? As it turns out, the answer to this question is positive. It is easy to show that some factorization terms coming from 4pt amplitude scale like ${\cal O}(\epsilon^2)$ while the sum scales like ${\cal O}(\epsilon^4)$.  For 6pt there is no need to add the contact term while at 8pt there is a single contact term, $\la12\ra^2\la34\ra^2[56]^2[78]^2s_{12}s_{34}s_{56}+$ perm., which we have to add. We can conjecture that for any $n$, the kinematic cancelations between Feynman diagrams giving ${\cal O}(\epsilon^4)$ behavior is very similar to BI theory but now individual graphs are decorated by particular $\Box=\partial^2$ acting on pairs of $-$ or $+$ legs. We obtain an infinite sequence of terms
\begin{equation}
{\cal L} = f + (f \Box f + g \Box g) + \Box^2 f^3 + \Box^2 fg^2 + \dots,
\end{equation}
where the term at  6pt order is only schematic form.
Note that this theory is not some generic higher derivative correction to the BI action since the improved ${\cal O}(\epsilon^4)$ behavior in the chiral multi-soft limit is not a consequence of SUSY. It remains to be seen if there is any physical significance to this theory and what extra amplitudes-level properties assumptions would specify it uniquely.

Finally, let us discuss how to constrain Eq.~(\ref{ansatz2}) from single soft limits via dimensional reduction.  Consider the case where all the vectors are dimensionally reduced to scalars which have the unique ${\cal O}(\epsilon^3)$ soft behavior of the special Galileon. Unlike DBI, the special Galileon does not have a multi-field analogue corresponding to multiple extra dimensions, so we are forced to dimensionally reduce all vectors to a single extra dimension, $\lbrace n\rbrace $.  This corresponds to the setting $(e_i\cdot p_j)=0$ and $(e_i\cdot e_j)=1$ for all indices $i,j$.  Perhaps unsurprisingly, this procedure yields {\it multiple} vector theories satisfying these constraints. For example, this is achieved by the Lagrangian,
\begin{multline}
{\cal L} = \sum_{n}c_n F^2 \,\epsilon^{\alpha_1\dots \alpha_D}\epsilon^{\beta_1\dots \beta_D} \partial_{\alpha_1}F_{\beta_1\alpha_D}\partial_{\beta_2}F_{\alpha_2\beta_D} \label{vGal} \\
\times\prod_{i=2}^n \partial_{\alpha_{2i-1}}F_{\beta_{2i-1}\mu_i} \partial_{\alpha_{2i}}F_{\beta_{2i}}^{\phantom{\beta_2}\mu_i} \prod_{j=2n+1}^{D-1}\hspace{-0.1cm} \eta_{\alpha_j\beta_j}.
\end{multline}
Under dimensional reduction this trivially reduces to a special Galileon in $d=D-1$ dimensions, 
\begin{align*}
\mathcal{L}=\sum_{n}c_n (\partial \phi)^2 \varepsilon ^{\mu _{1}\ldots \mu _{d}}\varepsilon ^{\nu
_{1}\ldots \nu _{d}}\prod_{k=1}^{2n}(\partial_{\mu_k}\partial _{\nu_k}\phi)
\hspace{-0.1cm}\prod\limits_{j=2n+1}^{d}\hspace{-0.1cm}\eta _{\mu _{j}\nu_{j}} ,
\end{align*}
where $c_n$ are certain combinatorial factors given in \cite{Hinterbichler:2015pqa}. Nevertheless, applying the simple replacement $\partial_{\alpha_k}F_{\beta_k\mu} \rightarrow\partial_{\beta_k}F_{\alpha_k\mu}$ to Eq.~(\ref{vGal}) yields a different, physically distinct vector Lagrangian whose dimensionally reduced scalar amplitudes are the same.  Hence the constraint of the soft limit and dimensional reduction into a single direction does not uniquely fix the amplitude. 

That said, imposing constraints from ${\cal O}(\epsilon^2)$ soft zeros for combinations of dimensional reduction actually fixes the 4pt amplitude uniquely from $\lbrace 2|2\rbrace $.  However, the 6pt amplitude still has free parameters after applying constraints from $\lbrace 4|2\rbrace $, $\lbrace 2|2|2\rbrace $, $\lbrace 4\rbrace $, $\lbrace 2|2\rbrace $ and $\lbrace 2\rbrace $.  So while this gives extra conditions, there are still not enough to fix the action completely. The question of whether there is a unique theory of this type given additional constraints is left for the future work.

\section{Conclusions}
In summary, we have applied modern amplitude methods to EFTs of massless vector particles. We have unambiguously identified BI theory as a theory uniquely fixed by certain infrared conditions.  This will hopefully serve as a novel tool for discovering new theoretical structures like the Galileon-like theories described here.

\medskip

\acknowledgments
{\it Acknowledgments:} 
We thank Nima Arkani-Hamed for useful discussions.
This work is supported in part by Czech Government projects GACR 18-17224S and LTAUSA17069. C.C.~and C.W.~are supported by a Sloan Research Fellowship and a DOE Early Career Award under grant no.~DE-SC0010255, J.T.~is supported by a DOE grant no.~DE-SC0009999.


\begin{thebibliography}{99}
\bibitem{note:8ptsoft}
More precisely, we find the 8pt amplitude behaves as ${\cal O}(\epsilon^2)$ if we send four spinors of the same type scale as $\epsilon$. However only ${\cal O}(\epsilon)$ behavior is required for the on-shell construction of the amplitude, thus this additional ${\cal O}(\epsilon^2)$ behavior shows a bonus relation. We have found similar ${\cal O}(\epsilon^2)$ at 10pt, and we will leave general analysis for the future work. If we send three of these four spinors to $\epsilon$, the amplitude goes as ${\cal O}(\epsilon)$.
\end{thebibliography}
\end{document}